\documentclass[aps,prd,preprint,showpacs,showkeys,superscriptaddress]{revtex4}

\usepackage{latexsym}
\usepackage{amsmath,amssymb}
\usepackage[dvips]{graphicx}
\usepackage[dvips]{hyperref} 

\begin{document}


\title{Note on nonsingular cyclic universes in the deformed Ho\v{r}ava-Lifshitz gravity}

\author{Edwin J. Son}
\email[]{eddy@sogang.ac.kr}
\affiliation{Center for Quantum Spacetime, Sogang University, Seoul 121-742, Korea}

\author{Wontae Kim}
\email[]{wtkim@sogang.ac.kr}
\affiliation{Center for Quantum Spacetime, Sogang University, Seoul 121-742, Korea}
\affiliation{Department of Physics, Sogang University, Seoul 121-742, Korea}

\date{\today}

\begin{abstract}
We perform the phase space analysis in terms of the linearization
technique in the Ho\v{r}ava-Lifshitz gravity with the \emph{softly}
broken detailed balance condition. It can be shown that the
bouncing universe appears only for the positive spatial curvature
of $k=+1$, and it is possible to obtain oscillating universe with
the help of the negative dark radiation and the negative
cosmological constant.
\end{abstract}



\maketitle



Recently, Ho\v{r}ava-Lifshitz (HL) gravity motivated by the Lifshitz theory
in the condensed matter physics~\cite{lifshitz} has been proposed as an
ultraviolet (UV) completion of general relativity~\cite{horava2,horava,horava3}.
The basic ingredient of the UV completion comes from an anisotropic scaling
between space and time,
\begin{equation}
\label{scaling}
t \to b^{z}\, t, \qquad x^i \to b\, x^i,
\end{equation}
where the conventional form can be recovered for the infrared (IR)
limit of $z =1$.
In the original HL gravity, the ``detailed balance'' condition (DBC) was introduced to simplify its quantum behavior and renormalization properties, and it has been shown that its renormalizability depends on the renormalizability of the topologically massive gravity (TMG)~\cite{djt,djt2} used for the DBC~\cite{or,or2}.
However, it is well-known that HL theory with the DBC can only admit solutions which are asymptotically AdS.
To solve this issue, Ho\v{r}ava considered a \emph{softly} broken model which contains IR modification without any change in UV terms, so that HL theory admits asymptotically flat solutions as well without loss of the renormalizability~\cite{horava}.

As pointed out by Ho\v{r}ava in Ref.~\cite{horava}, the \emph{softly} broken HL theory still satisfies the DBC at short distances, since the dominant terms (sixth order terms in spatial derivatives) are constructed by the Cotton-York tensor through the DBC.
The Cotton-York tensor can be derived by a variational principle with the gravitational Chern-Simons (GCS) action~\cite{djt,djt2}.
In other words, the renormalizability of the \emph{softly} broken HL theory in the UV limit, depends on the renormalizability of the GCS term.
Even though the subdominant terms (fourth order terms in spatial derivatives) are taken into account, they can be constructed from the TMG with the vanishing cosmological constant, so that its renormalizability depends on that of the TMG.
Therefore, it means that the present \emph{softly} broken HL theory is still renormalizable.

On the other hand, there has been much attention
to HL gravity to explore the cosmological implication of HL
theory~\cite{ts,kk,mukohyama,mnty,brandenberger,gwxb,ks,ww,mipark,cs,bps,ls,ces,wy,sj,jss,mmk} and related subject~\cite{ck,ac,cc,lmp,cy,gh}.
As an extension of HL theory, $F(R)$ HL gravity has been well appreciated to show that an accelerating phase of the universe can be obtained~\cite{cnoot,ccnoot}, and a unified description of inflation with dark energy is possible~\cite{enos,knos}.
In particular, bouncing universe has been intensively studied~\cite{wands,aw,fb,finelli,pp,pp2,bms,bbms,cqplz,cqbz} in general relativity (for a recent
review, see Refs.~\cite{brandenberger:review,brandenberger:review2}). 

In this note, we
would like to perform phase space analyses of the HL gravity with the \emph{softly} broken DBC. In fact, the original DBC severely restricts the asymptotic behaviors so that the only asymptotically anti-de Sitter spacetime is possible. What it means is that at the late-time stage of the universe the decelerated expansion may appear, unfortunately, which seems to be incompatible with the recent accelerated expansion of the universe. By the way, it is possible to obtain the asymptotically flat solution of the vanishing cosmological constant along with the asymptotically anti-de Sitter solution for the \emph{softly} broken DBC. So, one may ask what the crucial cosmological difference is between the case of the vanishing cosmological constant and the nonvanishing one in the \emph{softly} broken DBC model,
which is a main motivation of the present work. 
Eventually, it turns out that the cosmological behaviors between them are not so much different from each other. It shows that only for $k=+1$, cyclic universes exist throughout matter bounce at early stage and decelerated expansion at late time for both cases.
As for $k =-1$, there exists an unstable fixed point in the forbidden region of the negative total energy for both cases. In this case, we can avoid the initial singularity by imposing the positivity of the total energy. 
Unfortunately, all of them do not show the recent accelerated expansion, which is a weakness of the model with the DBC either with the \emph{softly} broken DBC. 


Now, considering Arnowitt-Deser-Misner (ADM) decomposition of the
metric with $ds^2 = - N^2 c^2 dt^2 + g_{ij} (dx^i + N^i dt) (dx^j
+ N^j dt)$~\cite{adm,adm2}, the Einstein-Hilbert action can be
rewritten as
\begin{equation}
\label{act:EH}
\begin{aligned}
I_{EH} &= \frac{c^3}{16\pi G_N} \int d^4x \sqrt{-\mathcal{G}} \left[ \mathcal{R} - 2\Lambda \right] \\
  &= \frac{c^2}{16\pi G_N} \int dt d^3x \sqrt{g} N \left[ K_{ij} K^{ij} - K^2 + c^2 \left( R - 2\Lambda \right) \right],
\end{aligned}
\end{equation}
where $K_{ij} \equiv \frac{1}{2N} \left[ \dot{g}_{ij} - \nabla_i
N_j - \nabla_j N_i \right]$ is the extrinsic curvature at $t=
\text{constant}$ hyper-surface, and the dot denotes the derivative
with respect to time $t$. Here, $g_{ij}$, $R$, and $\nabla_i$ are the
metric, the intrinsic curvature, and the covariant derivative in
the three-dimensional hyper-surface, respectively. The above
anisotropic scaling~\eqref{scaling} gives
$N^i \to b^{1-z} N^i$ and $c \to b^{1-z} c$ while $g_{ij}$ and $N$
are invariant. Requiring that the Planck constant is also
invariant, $\hbar\to\hbar$, we get $E\to b^{-z}E$ and $M\to
b^{z-2}M$, where $E$ and $M$ are energy and mass, respectively.
Then, the Ho\v{r}ava-Lifshitz gravity (HL) under the \emph{softly}
broken DBC~\cite{horava} is given by
\begin{equation}
\label{action}
\begin{aligned}
I_\text{HL} = \int dt d^3x \sqrt{g} N \bigg[ & \frac{2}{\kappa^2} \left( K_{ij} K^{ij} - \lambda K^2 \right) - \frac{\kappa^2}{2\zeta^4} C_{ij} C^{ij} + \frac{\kappa^2\mu}{2\zeta^2} \varepsilon^{ijk} R_{i\ell} \nabla_j R_k^\ell \\
  &- \frac{\kappa^2\mu^2}{8} R_{ij} R^{ij} + \frac{\kappa^2\mu^2}{8(3\lambda-1)} \left( \frac{4\lambda-1}{4} R^2 + ( \omega - \Lambda_W ) R + 3 \Lambda_W^2 \right)\bigg],
\end{aligned}
\end{equation}
where $\kappa^2$ is a coupling related to the Newton constant
$G_N$, and $\lambda$ is an additional dimensionless coupling
constant. In addition, $\omega$ represents the IR
modification which is essential to have asymptotically flat
solutions. The coupling constants $\mu$, $\Lambda_W$, and $\zeta$
come from the three-dimensional Euclidean topologically massive
gravity action~\cite{djt,djt2} given by
\begin{equation}
W = \mu \int d^3x \sqrt{g} (R-2\Lambda_W) + \frac{1}{\zeta^2} \int \chi(\Gamma),
\end{equation}
where $\chi(\Gamma)$ represents the gravitational Chern-Simons term.
Then, the scaling of couplings can be obtained as $\kappa^2\to b^{3-z}\kappa^2$, $\mu\to b^{-1}\mu$, $\Lambda_W\to b^{-2}\Lambda_W$, $\zeta\to\zeta$ and $\omega\to b^{-2}\omega$.
Note that identifying the fundamental constants with
\begin{equation}
\label{fund:const}
c = \frac{\kappa^2}{4} \sqrt{\frac{\mu^2(\omega-\Lambda_W)}{3\lambda-1}}, \qquad G_N = \frac{\kappa^2c^2}{32\pi}, \qquad \Lambda = -\frac{3\Lambda_W^2}{2(\omega-\Lambda_W)},
\end{equation}
the Einstein-Hilbert action~\eqref{act:EH} can be recovered in the IR limit when $\lambda=1$.
For $\lambda>1/3$, we have $\omega>\Lambda_W$ and $\Lambda<0$ from the identification~\eqref{fund:const}, while for $\lambda<1/3$, we have $\omega<\Lambda_W$ and $\Lambda>0$.
From now on, we will assume $\lambda>1/3$ to keep the Einstein limit.

Now, we are going to consider the Robertson-Walker (RW) metric,
\begin{equation}
\label{met:RW}
ds^2 = - c^2 dt^2 + a^2(t) \left[ \frac{dr^2}{1-kr^2} + r^2 d\Omega_2^2 \right],
\end{equation}
where $k=0,\pm1$ are normalized spatial curvature. After some
tedious calculations, the equations of motion from~\eqref{action}
can be obtained as
\begin{align}
3 (3\lambda-1) H^2 
  =&\ \frac{\kappa^2}{2} \rho + 6 c^2 \left[ \frac{\Lambda}{3} - \frac{k}{a^2} - \frac{k^2}{2a^4(\omega-\Lambda_W)} \right], \notag \\
  =&\ \frac{\kappa^2}{2} \left[ \rho + \rho_\text{vac} + \rho_k + \rho_\text{dr} \right], \label{eq:energy} \\
(3\lambda-1) \left( \dot{H} + \frac32 H^2 \right) 
  =&\ -\frac{\kappa^2}{4} p - 3c^2 \left[ -\frac{\Lambda}{3} + \frac{k}{3a^2} - \frac{k^2}{6a^4(\omega-\Lambda_W)} \right], \notag \\
  =&\ -\frac{\kappa^2}{4} \left[ p + p_\text{vac} + p_k + p_\text{dr} \right], \label{eq:pressure}
\end{align}
where $H=\dot{a}/a $ is the Hubble parameter.
Apart from the normal matter contributions of $\rho$ and $p$,
the \emph{additional} energy-momentum contributions from the potential
terms~\eqref{action} are
explicitly written as
\begin{equation}
\label{add:en}
  p_\text{vac} = -\rho_\text{vac} = -
  \frac{12c^2}{\kappa^2} \frac{\Lambda}{3},
    \quad 
    p_k = -\frac13\rho_k = 
    \frac{4c^2}{\kappa^2} \frac{k}{a^2}, 
  \quad
  p_\text{dr} = \frac13\rho_\text{dr} = -
  \frac{4c^2}{\kappa^2} \frac{k^2}{2a^4(\omega-\Lambda_W)}.
\end{equation}
Similarly to the general relativity, $\rho_\text{vac}$
and $\rho_k$
come from the vacuum energy and the spatial curvature
contributions, respectively. In particular, in the HL theory,
$\rho_\text{dr}$
is called the \emph{dark radiation}.
The total \emph{actual} energy can be defined by
$\rho_\text{tot}=\rho + \rho_\text{vac}  + \rho_\text{dr}$.
Note that the negative dark radiation is enough
to realize the matter bounce~\cite{brandenberger,gwxb}, so we do not need
any additional \emph{ad hoc} energy sources to make the bouncing
universe.
By the way, combining Eqs.~\eqref{eq:energy} and~\eqref{eq:pressure},
we can get the
acceleration of the scale factor,
\begin{equation}
\label{eq:acc}
(3\lambda-1) \frac{\ddot{a}}{a} = -\frac{\kappa^2}{12} \left[ \rho_\text{tot} + 3 p_\text{tot} \right]
= -\frac{\kappa^2}{12} [1+3\omega(a)] \rho_\text{tot},
\end{equation}
where we used the relation $\rho_k+3p_k=0$ and $\omega(a) =  p_\text{tot}/\rho_\text{tot}$ is the equation-of-state parameter for the total energy.



Now, let us perform numerical analysis for this system~\eqref{eq:acc} and obtain phase portraits. First of all, we should
introduce additional variable $v=\dot{a}$, so that
we  get two first-order differential equations,
\begin{align}
& \dot{a} = v, \label{eq:a} \\ 
& \dot{v} = -\frac{1}{2a} \left[ 1 + 3 \omega(a) \right] \left( v^2 + \frac{2c^2k}{3\lambda-1} \right), \label{eq:v} 
\end{align}
where we used Eqs.~\eqref{eq:energy} and \eqref{eq:acc}. 
Next, following the linearization technique~\cite{strogatz}, the Jacobian matrix is obtained as
\begin{equation}
\label{mat:A}
A = \left[
\begin{array}{cc}
  0 & 1 \\
  -3c^2k\omega'(a^*)/(3\lambda-1)a^* & 0
\end{array}
\right]
\end{equation}
and the corresponding eigenvalues are
$\eta=\pm\sqrt{-3c^2k\omega'(a^*)/(3\lambda-1)a^*}$,
then fixed points $(a^*,0)$ are classified as a center for $k\omega'(a^*)>0$ or a saddle node for $k\omega'(a^*)<0$, where $a^*$ satisfies $1+3\omega(a^*)=0$.

We are now in a position to specify the matter source which consists of
the conventional cold matter and radiation, $\rho=\rho_m+\rho_r$
with $\rho_m\sim a^{-3}$ and $\rho_r\sim a^{-4}$.
Then, the equation-of-state parameter $\omega(a)$ is given by
\begin{equation}
\label{eos}
\omega(a) = \frac{\Omega_\text{dr}+\Omega_r-3a^4\Omega_\text{vac}}{3 \left[ \Omega_\text{dr}+\Omega_r+a\Omega_{m}+a^4\Omega_\text{vac} \right]},
\end{equation}
where the density parameters are defined as $\Omega_i=\rho_i/\rho_c$
evaluated at the present universe scale $a_0$, which can be fixed to
$a_0=1$ for convenience.
\begin{figure}[pbt]
  \includegraphics[width=0.45\textwidth]{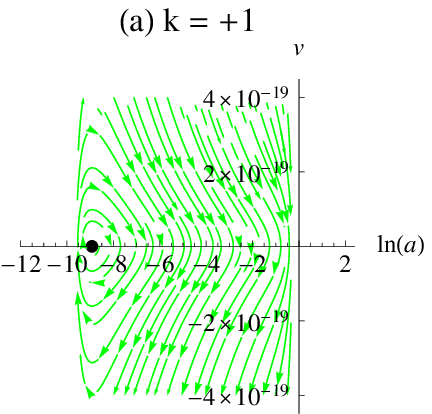}
  \hspace{0.05\textwidth}
 \includegraphics[width=0.45\textwidth]{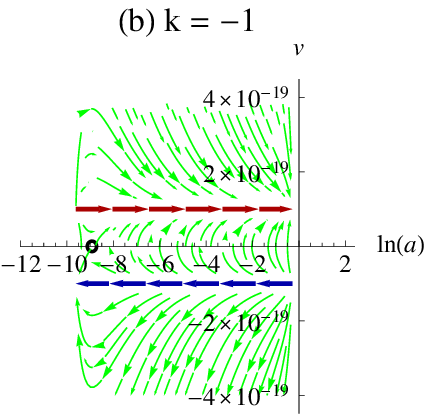}
  \caption{\label{fig:pp:inf}
  The phase portraits are plotted in the
  logarithmic scale for (a) $k=+1$ with the stable fixed point and (b) $k=-1$ with the unstable fixed point.  The black and white dots on the $a$-axes represent fixed points $(a^*,0)$ satisfying $1+3\omega(a^*)=0$.}
\end{figure}
Assigning $\Omega_\text{vac}\simeq-0.7$, $\Omega_m\simeq0.3$,
$\Omega_k\sim-10^{-2}k$, $\Omega_r\sim10^{-4}$, and
$\Omega_\text{dr}\sim-1.2\times10^{-4}$, 
the phase
portraits for this system are plotted in Fig.~\ref{fig:pp:inf} in
the logarithmic scale for (a) $k=+1$ and (b) $k=-1$, respectively.
The black and white dots on the $a$-axes represent fixed points.
It is of interest to show that in Fig.~\ref{fig:pp:inf} the
bouncing universe appears only for the positive spatial curvature
of $k=+1$, and it is possible to obtain oscillating universe with
the help of the negative dark radiation and the negative
cosmological constant. We have maintained the positivity of the
total energy density. In this respect, the inside between the
horizontal arrows in Fig.~\ref{fig:pp:inf} (b) is not allowed. As
for the negative contribution of dark radiation, it is interesting
to note that there have been concrete justifications for models
with negative density, in particular, a brane universe moving in a
curved higher dimensional bulk space~\cite{kk:mirage} and a model
of dark energy stemming from a fermionic condensate~\cite{abc},
although we do not know how to justify the negative dark radiation
in this model.
\begin{figure}[pbt]
  \includegraphics[width=0.45\textwidth]{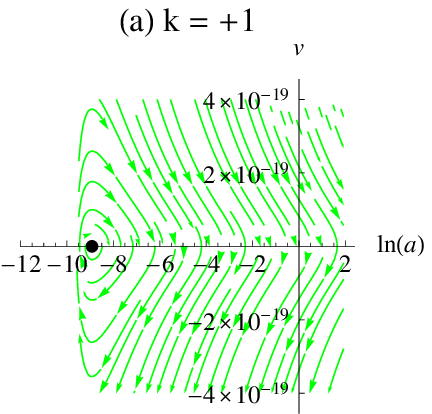}
  \hspace{0.05\textwidth}
 \includegraphics[width=0.45\textwidth]{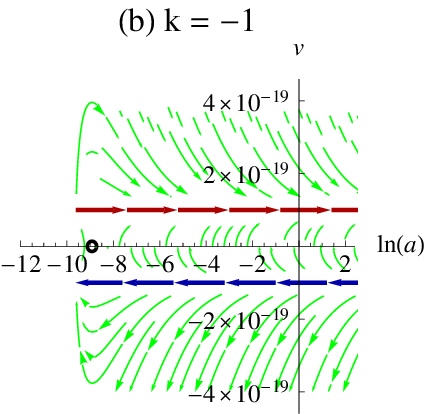}
  \caption{\label{fig:pp:inf0}
  The phase portraits with same data as in Fig.~\ref{fig:pp:inf} except for vanishing cosmological constant are plotted in the
  logarithmic scale for (a) $k=+1$ with the stable fixed point and (b) $k=-1$ with the unstable fixed point.  The black and white dots on the $a$-axes represent fixed points $(a^*,0)$ satisfying $1+3\omega(a^*)=0$.}
\end{figure}
In Fig.~\ref{fig:pp:inf0}, the cosmological evolution can be shown for the case of the vanishing cosmological constant following the same procedure.
Eventually, it turns out that the cosmological behaviors between the cases of the vanishing cosmological constant (Fig.~\ref{fig:pp:inf0}) and the nonvanishing cosmological constant (Fig.~\ref{fig:pp:inf}) are not so much different from each other.

Final comment is in order. Unfortunately, the HL cosmology with
the DBC does not exhibit the recent accelerated expansion of our
universe. On the other hand, a positive cosmological constant can
be obtained for $\lambda>1/3$ by an analytic continuation, $\mu\to
i\mu$ and $\zeta^2\to-i\zeta^2$~\cite{lmp,cy}, which makes the
coefficients of the potential terms in the action~\eqref{action}
have the opposite sign, so that we have $\omega<\Lambda_W$ and
$\Lambda>0$. Then, it gives the late-time accelerated expansion,
however, it causes the initial singularity problem because of the
absence of the matter bounce.


\begin{acknowledgments}
E. Son was supported by the National Research Foundation of
Korea(NRF) grant funded by the Korea government(MEST) through the
Center for Quantum Spacetime(CQUeST) of Sogang University with grant
number 2005-0049409. W. Kim was supported
by the Special Research Grant of Sogang University, 200911044.
\end{acknowledgments}



\begin{thebibliography}{99}

\bibitem{lifshitz}
  E.~M.~Lifshitz,
  Zh.\ Eksp.\ Teor.\ Fiz.\  {\bf 11}, 255 (1941).
%

\bibitem{horava2}
  P.~Ho\v{r}ava,
%
  JHEP {\bf 0903}, 020 (2009)
  [\texttt{arXiv:0812.4287 [hep-th]}].

\bibitem{horava}
  P.~Ho\v{r}ava,
  Phys.\ Rev.\  D {\bf 79}, 084008 (2009)
  [\texttt{arXiv:0901.3775 [hep-th]}].

\bibitem{horava3}
  P.~Ho\v{r}ava,
  Phys.\ Rev.\ Lett.\  {\bf 102}, 161301 (2009)
  [\texttt{arXiv:0902.3657 [hep-th]}].


\bibitem{djt}
  S.~Deser, R.~Jackiw and S.~Templeton,
  Annals Phys.\  {\bf 140}, 372 (1982)
  [Erratum-ibid.\  {\bf 185}, 406 (1988); ibid.\ {\bf 281}, 409-449 (2000)].

\bibitem{djt2}
  S.~Deser, R.~Jackiw and S.~Templeton,
  Phys.\ Rev.\ Lett.\  {\bf 48}, 975 (1982).


\bibitem{or}
  D.~Orlando and S.~Reffert,
  Class.\ Quant.\ Grav.\  {\bf 26}, 155021 (2009)
  [\texttt{arXiv:0905.0301 [hep-th]}].

\bibitem{or2}
  D.~Orlando and S.~Reffert,
  Phys.\ Lett.\  B {\bf 683}, 62 (2010)
  [\texttt{arXiv:0908.4429 [hep-th]}].


\bibitem{ts}
  T.~Takahashi and J.~Soda,
  Phys.\ Rev.\ Lett.\  {\bf 102}, 231301 (2009)
  [\texttt{arXiv:0904.0554 [hep-th]}].

\bibitem{kk}
  E.~Kiritsis and G.~Kofinas,
  Nucl.\ Phys.\  B {\bf 821}, 467 (2009)
  [\texttt{arXiv:0904.1334 [hep-th]}].

\bibitem{mukohyama}
  S.~Mukohyama,
  JCAP {\bf 0906}, 001 (2009)
  [\texttt{arXiv:0904.2190 [hep-th]}].

\bibitem{mnty}
  S.~Mukohyama, K.~Nakayama, F.~Takahashi and S.~Yokoyama,
  Phys.\ Lett.\  B {\bf 679}, 6 (2009)
  [\texttt{arXiv:0905.0055 [hep-th]}].

\bibitem{brandenberger}
  R.~Brandenberger,
  Phys.\ Rev.\  D {\bf 80}, 043516 (2009)
  [\texttt{arXiv:0904.2835 [hep-th]}];

\bibitem{gwxb}
  X.~Gao, Y.~Wang, W.~Xue and R.~Brandenberger,
  JCAP {\bf 1002}, 020 (2010).
  [\texttt{arXiv:0911.3196 [hep-th]}].

\bibitem{ks}
  A.~Kehagias and K.~Sfetsos,
  Phys.\ Lett.\  B {\bf 678}, 123 (2009)
  [\texttt{arXiv:0905.0477 [hep-th]}].

\bibitem{ww}
  A.~Wang and Y.~Wu,
  JCAP {\bf 0907}, 012 (2009)
  [\texttt{arXiv:0905.4117 [hep-th]}].

\bibitem{mipark}
  M.-I.~Park,
  JHEP {\bf 0909}, 123 (2009)
  [\texttt{arXiv:0905.4480 [hep-th]}].

\bibitem{cs}
  Y.~F.~Cai and E.~N.~Saridakis,
  JCAP {\bf 0910}, 020 (2009)
  [\texttt{arXiv:0906.1789 [hep-th]}].

\bibitem{bps}
  D.~Blas, O.~Pujolas and S.~Sibiryakov,
  Phys.\ Rev.\ Lett.\  {\bf 104}, 181302 (2010)
  [\texttt{arXiv:0909.3525 [hep-th]}].

\bibitem{ls}
  G.~Leon and E.~N.~Saridakis,
  JCAP {\bf 0911}, 006 (2009)
  [\texttt{arXiv:0909.3571 [hep-th]}].

\bibitem{ces}
  S.~Carloni, E.~Elizalde and P.~J.~Silva,
  Class.\ Quant.\ Grav.\  {\bf 27}, 045004 (2010)
  [\texttt{arXiv:0909.2219 [hep-th]}].

\bibitem{wy}
  P.~Wu and H.~W.~Yu,
  Phys.\ Rev.\  D {\bf 81}, 103522 (2010)
  [\texttt{arXiv:0909.2821 [gr-qc]}].

\bibitem{sj}
  M.~R.~Setare and M.~Jamil,
  JCAP {\bf 1002}, 010 (2010)
  [\texttt{arXiv:1001.1251 [hep-th]}].

\bibitem{jss}
  M.~Jamil, E.~N.~Saridakis and M.~R.~Setare,
  \textit{The generalized second law of thermodynamics in Horava-Lifshitz
  cosmology},
  \texttt{arXiv:1003.0876 [hep-th]}.

\bibitem{mmk}
  K.~i.~Maeda, Y.~Misonoh and T.~Kobayashi,
  Phys.\ Rev.\  D {\bf 82}, 064024 (2010)
  [\texttt{arXiv:1006.2739 [hep-th]}].


\bibitem{lmp}
  H.~Lu, J.~Mei and C.~N.~Pope,
  Phys.\ Rev.\ Lett.\  {\bf 103}, 091301 (2009)
  [\texttt{arXiv:0904.1595 [hep-th]}].

\bibitem{cy}
  E.~O Colgain and H.~Yavartanoo,
  JHEP {\bf 0908}, 021 (2009)
  [\texttt{arXiv:0904.4357 [hep-th]}].


\bibitem{ck}
  R.~G.~Cai and S.~P.~Kim,
  JHEP {\bf 0502}, 050 (2005)
  [\texttt{arXiv:hep-th/0501055}].

\bibitem{ac}
  M.~Akbar and R.~G.~Cai,
  Phys.\ Rev.\  D {\bf 75}, 084003 (2007)
  [\texttt{arXiv:hep-th/0609128}].

\bibitem{cc}
  R.~G.~Cai and L.~M.~Cao,
  Phys.\ Rev.\  D {\bf 75}, 064008 (2007)
  [\texttt{arXiv:gr-qc/0611071}].

\bibitem{gh}
  A.~Ghodsi and E.~Hatefi,
  Phys.\ Rev.\  D {\bf 81}, 044016 (2010)
  [\texttt{arXiv:0906.1237 [hep-th]}].


\bibitem{cnoot}
  M.~Chaichian, S.~Nojiri, S.~D.~Odintsov, M.~Oksanen and A.~Tureanu,
  Class.\ Quant.\ Grav.\  {\bf 27}, 185021 (2010)
  [\texttt{arXiv:1001.4102 [hep-th]}];

\bibitem{ccnoot}
  S.~Carloni, M.~Chaichian, S.~Nojiri, S.~D.~Odintsov, M.~Oksanen and A.~Tureanu,
  Phys.\ Rev.\  D {\bf 82}, 065020 (2010)
  [\texttt{arXiv:1003.3925 [hep-th]}].

\bibitem{enos}
  E.~Elizalde, S.~Nojiri, S.~D.~Odintsov and D.~Saez-Gomez,
  Eur.\ Phys.\ J.\ C {\bf 70}, 351 (2010)
  [\texttt{arXiv:1006.3387 [hep-th]}];

\bibitem{knos}
  J.~Kluson, S.~Nojiri, S.~D.~Odintsov and D.~Saez-Gomez,
  \textit{U(1) Invariant F(R) Horava-Lifshitz Gravity},
  \texttt{arXiv:1012.0473 [hep-th]}.


\bibitem{wands}
  D.~Wands,
  Phys.\ Rev.\  D {\bf 60}, 023507 (1999)
  [\texttt{gr-qc/9809062}].

\bibitem{aw}
  L.~E.~Allen and D.~Wands,
  Phys.\ Rev.\  D {\bf 70}, 063515 (2004)
  [\texttt{astro-ph/0404441}].

\bibitem{fb}
  F.~Finelli and R.~Brandenberger,
  Phys.\ Rev.\  D {\bf 65}, 103522 (2002)
  [\texttt{hep-th/0112249}].

\bibitem{finelli}
  F.~Finelli,
  JCAP {\bf 0310}, 011 (2003)
  [\texttt{hep-th/0307068}].

\bibitem{pp}
  P.~Peter and N.~Pinto-Neto,
  Phys.\ Rev.\  D {\bf 66}, 063509 (2002)
  [\texttt{hep-th/0203013}].

\bibitem{pp2}
  P.~Peter and N.~Pinto-Neto,
  Phys.\ Rev.\  D {\bf 78}, 063506 (2008)
  [\texttt{arXiv:0809.2022 [gr-qc]}].

\bibitem{bms}
  T.~Biswas, A.~Mazumdar and W.~Siegel,
  JCAP {\bf 0603}, 009 (2006)
  [\texttt{hep-th/0508194}].

\bibitem{bbms}
  T.~Biswas, R.~Brandenberger, A.~Mazumdar and W.~Siegel,
  JCAP {\bf 0712}, 011 (2007)
  [\texttt{hep-th/0610274}].

\bibitem{cqplz}
  Y.~F.~Cai, T.~Qiu, Y.~S.~Piao, M.~Li and X.~Zhang,
  JHEP {\bf 0710}, 071 (2007)
  [\texttt{arXiv:0704.1090 [gr-qc]}].

\bibitem{cqbz}
  Y.~F.~Cai, T.~Qiu, R.~Brandenberger and X.~Zhang,
  Phys.\ Rev.\  D {\bf 80}, 023511 (2009)
  [\texttt{arXiv:0810.4677 [hep-th]}].


\bibitem{brandenberger:review}
  R.~H.~Brandenberger,
  \textit{Alternatives to Cosmological Inflation},
  \texttt{arXiv:0902.4731 [hep-th]}.

\bibitem{brandenberger:review2}
  R.~H.~Brandenberger,
  \textit{Cosmology of the Very Early Universe},
  \texttt{arXiv:1003.1745 [hep-th]}.


\bibitem{adm}
  R.~L.~Arnowitt, S.~Deser and C.~W.~Misner,
  Phys.\ Rev.\  {\bf 117}, 1595 (1960).

\bibitem{adm2}
  R.~L.~Arnowitt, S.~Deser and C.~W.~Misner,
  ``The dynamics of general relativity''
  in \textit{Gravitation: an introduction to current research}, ed. L.~Witten pp 227-265 (New York: Wiley 1962)
  [\texttt{gr-qc/0405109}].


\bibitem{strogatz}
  S.~H.~Strogatz,
  ``Nonlinear Dynamics and Chaos. With Applications to Physics, Biology, Chemistry, and Engineering,''
{\it  Westview Press (2000) 498 p}.


\bibitem{kk:mirage}
  A.~Kehagias and E.~Kiritsis,
  JHEP {\bf 9911}, 022 (1999)
  [\texttt{arXiv:hep-th/9910174}].

\bibitem{abc}
  S.~Alexander, T.~Biswas and G.~Calcagni,
  Phys.\ Rev.\  D {\bf 81}, 043511 (2010)
  [\texttt{arXiv:0906.5161 [astro-ph.CO]}].

\end{thebibliography}
\end{document}